\documentstyle[12pt]{article}

\font\twlgot =eufm10 scaled \magstep1 \font\egtgot =eufm8
\font\sevgot =eufm7 \font\twlmsb =msbm10 scaled \magstep1
\font\egtmsb =msbm8 \font\sevmsb =msbm7

\newfam\gotfam

\textfont\gotfam\twlgot \scriptfont\gotfam\egtgot
\scriptscriptfont\gotfam\sevgot

\newfam\msbfam
\textfont\msbfam\twlmsb \scriptfont\msbfam\egtmsb
\scriptscriptfont\msbfam\sevmsb
\def\Bbb{\protect\pBbb}
\def\pBbb{\relax\ifmmode\expandafter\Bb\else\typeout{You cann't use
Bbb in text mode}\fi}
\def\Bb #1{{\fam\msbfam\relax#1}}

\def\thebibliography#1{\bigskip\section*{}\bigskip\list
{$^{\arabic{enumi}}$}{\settowidth\labelwidth{#1}\leftmargin\labelwidth
\advance\leftmargin\labelsep
\usecounter{enumi}}
\def\newblock{\hskip .11em plus .33em minus .07em}
\sloppy\clubpenalty4000\widowpenalty4000 \sfcode`\.=1000\relax}

\def\op#1{\mathop{\fam0 #1}\limits}

\newcommand{\im}{{\rm Im\,}}

\newcommand{\beq}{\begin{equation}}
\newcommand{\eeq}{\end{equation}}
\newcommand{\ben}{\begin{eqnarray}}
\newcommand{\een}{\end{eqnarray}}
\newcommand{\be}{\begin{eqnarray*}}
\newcommand{\ee}{\end{eqnarray*}}
\newcommand{\bea}{\begin{eqalph}}
\newcommand{\eea}{\end{eqalph}}
\newcommand{\cA}{{\cal A}}

\newcommand{\cP}{{\cal P}}

\newcommand{\cL}{{\cal L}}
\newcommand{\cJ}{{\cal J}}

\newcommand{\cE}{{\cal E}}

\newcommand{\cS}{{\cal S}}
\newcommand{\cC}{{\cal C}}

\newcommand{\cG}{{\cal G}}
\newcommand{\bL}{{\bf L}}

\newcommand{\dl}{\delta}
\newcommand{\la}{\lambda}
\newcommand{\La}{\Lambda}
\newcommand{\f}{\phi}
\newcommand{\om}{\omega}

\newcommand{\m}{\mu}

\newcommand{\th}{\theta}

\newcommand{\vt}{\vartheta}

\newcommand{\up}{\upsilon}

\newcommand{\si}{\sigma}
\newcommand{\Si}{\Sigma}
\newcommand{\w}{\wedge}

\newcommand{\ol}{\overline}
\newcommand{\dr}{\partial}
\newcommand{\ar}{\op\longrightarrow}
\newcommand{\llr}{\op\longleftarrow}
\newcommand{\lto}{\leftarrow}
\newcommand{\ot}{\otimes}
\newcommand{\ap}{\approx}

\newcommand{\rdr}{\stackrel{\leftarrow}{\dr}{}}

\let\ssection=\section
\renewcommand{\section}{\setcounter{equation}{0}\ssection}

\newcounter{eqalph}
\newcounter{equationa}
\newcounter{remark}
\newcounter{example}
\newcounter{theorem}
\newcounter{proposition}
\newcounter{lemma}
\newcounter{corollary}
\newcounter{definition}
\newenvironment{eqalph}{\stepcounter{equation}
\setcounter{equationa}{\value{equation}} \setcounter{equation}{0}

\begin{eqnarray}}{\end{eqnarray}\setcounter{equation}{\value{equationa}}}

\def\theremark{\arabic{remark}}

\def\thetheorem{\arabic{theorem}}

\newenvironment{proof}{
{\it Proof:}}{}

\newenvironment{theo}{\refstepcounter{theorem}
{\bf Theorem \thetheorem:}}{}

\newenvironment{lem}{\refstepcounter{theorem}
{\bf Lemma \thetheorem:}}{}

\newcommand{\mar}[1]{}

\begin{document}
\hbox{}

{\parindent=0pt

{\large\bf Gauge conservation laws in a general setting.
Superpotential}
\bigskip

{\sc G. Sardanashvily}

{\sl Department of Theoretical Physics, Moscow State University,
117234 Moscow, Russia}

\bigskip
\bigskip

The fact that the conserved current of a gauge symmetry is reduced
to a superpotential is proved in a very general setting.
 }

\bigskip
\bigskip

\noindent {\bf I. INTRODUCTION}
\bigskip

The fact that the conserved current of a gauge symmetry is reduced
to a superpotential has been stated in different particular
variants, e.g., gauge theory of principal connections and gauge
gravitation theory.$^{1-4}$ We aim to prove this assertion in a
very general setting.

Generic higher-order Lagrangian theory of even and odd fields on
an $n$-dimensional smooth manifold $X$ and its variational
generalized supersymmetries (henceforth symmetries) are
considered.$^{5-8}$ These symmetries form a real vector space
$\cG_L$. In a general setting, a gauge symmetry of a Lagrangian
$L$ is defined as a $\cG_L$-valued linear differential operator on
some Grassmann-graded projective $C^\infty(X)$-module of finite
rank.$^{7,8}$ Note that any Lagrangian possesses gauge symmetries
which therefore must be separated into the trivial and non-trivial
ones. However, there is a problem of defining non-trivial gauge
symmetries.$^7$

In contrast with gauge symmetries, non-trivial Noether identities
of Lagrangian field theory are well described in homology
terms.$^{6-8}$ Therefore, we define non-trivial gauge symmetries
as those associated to complete non-trivial Noether identities in
accordance with the second Noether theorem (Theorem \ref{s4}).

Given a non-trivial gauge symmetry of a Lagrangian $L$, the
corresponding current $\cJ$ (\ref{09200}) is conserved by virtue
of the first Noether theorem (Theorem \ref{j45}). We prove that
this current takes the superpotential form
\be
\cJ^\m= W^\m + d_\nu U^{\nu\m}
\ee
where the term $W^\m$ vanishes on the kernel of the
Euler--Lagrange operator $\dl L$ (\ref{0709'}) of $L$ and
$U^{\nu\m}=-U^{\m\nu}$ is a superpotential (Theorem \ref{supp}).

\bigskip
\bigskip

\noindent {\bf II. LAGRANGIAN THEORY OF EVEN AND ODD FIELDS}
\bigskip

Lagrangian theory of even and odd fields is adequately formulated
in terms of the Grassmann-graded variational bicomplex on fiber
bundles and graded manifolds.$^{5,8,9}$ In a very general setting,
let us consider a composite bundle $F\to Y\to X$ where $F\to Y$ is
a vector bundle provided with bundle coordinates $(x^\la, y^i,
q^a)$. The jet manifolds $J^rF$ of $F\to X$ also are vector
bundles $J^rF\to J^rY$ coordinated by $(x^\la, y^i_\La, q^a_\La)$,
$0\leq |\La|\leq r$, where $\La=(\la_1...\la_k)$, $|\La|=k$,
denotes a symmetric multi-index. Let $(J^rY,\cA_r)$ be a graded
manifold whose body is $J^rY$ and whose structure ring $\cA_r$ of
graded functions consists of sections of the exterior bundle
\be
\w (J^rF)^*=\Bbb R\op\oplus (J^rF)^*\oplus\op\w^2
(J^rF)^*\op\oplus\cdots,
\ee
where $(J^rF)^*$ is the dual of $J^rF\to J^rY$. The local odd
basis for this ring is $\{c^a_\La\}$, $0\leq |\La|\leq r$. Let
$\cS^*_r[F;Y]$ be the differential graded algebra (henceforth DGA)
of graded differential forms on the graded manifold
$(J^rY,\cA_r)$. The inverse system of jet manifolds $J^{r-1}Y
\leftarrow J^rY$ yields the direct system of DGAs
\be
\cS^*[F;Y]\ar \cS^*_1[F;Y]\ar\cdots \cS^*_r[F;Y]\ar\cdots.
\ee
Its direct limit $\cS^*_\infty[F;Y]$ is the DGA of all graded
differential forms on graded manifolds $(J^rY,\cA_r)$. It is a
$C^\infty(Y)$-algebra locally generated by elements
$(c^a_\La,dx^\la,dy^i_\La, dc^a_\La)$, $0\leq |\La|$. Let us
recall the formulas
\be
\f\w\f' =(-1)^{|\f||\f'| +[\f][\f']}\f'\w \f, \qquad d(\f\w\f')=
d\f\w\f' +(-1)^{|\f|}\f\w d\f,
\ee
where $[\f]$ denotes the Grassmann parity. The collective symbol
$(s^A)$ further stands for the tuple $(y^i,c^a)$, called the local
basis for the DGA $\cS^*_\infty[F;Y]$. Let us denote
$[A]=[s^A]=[s^A_\La]$.

The DGA $\cS^*_\infty[F;Y]$ is split into the Grassmann-graded
variational bicomplex of $\cS^0_\infty[F;Y]$-modules
$\cS^{k,r}_\infty[F;Y]$ of $r$-horizontal and $k$-contact graded
forms locally generated by the one-forms $dx^\la$ and
$\th^A_\La=ds_\La^A -s^A_{\la +\La} dx^\la$. This bicomplex
contains the variational subcomplex
\mar{mar000}\beq
0\to \Bbb R\ar \cS^0_\infty[F;Y]\ar^{d_H}\cS^{0,1}_\infty[F;Y]
\cdots \ar^{d_H} \cS^{0,n}_\infty[F;Y]\ar^\dl
\cS^{1,n}_\infty[F;Y], \label{000}
\eeq
whose coboundary operator
\be
&& d_H(\f)=dx^\la\w d_\la\f= dx^\la\w(\dr_\la +
\op\sum_{0\leq|\La|} s_{\la\La}^A\dr^\La_A)\f,\\
&& d_H\circ h_0=h_0\circ d, \qquad h_0(\th^A_\La)=0, \qquad
h_0(dx^\la)=dx^\la,
\ee
is the total differential, and whose elements
\mar{0709,'}\ben
&& L=\cL \om \in \cS^{0,n}_\infty[F;Y], \qquad \om=dx^1\w\cdots\w dx^n, \label{0709}\\
&&  \dl L= \th^A\w \cE_A \om=\op\sum_{0\leq|\La|}
(-1)^{|\La|}\th^A\w d_\La (\dr^\La_A \cL) \om,  \qquad
d_\La=d_{\la_1}\cdots d_{\la_k}, \label{0709'}
\een
are graded Lagrangians and their Euler--Lagrange operators.
Further, a pair $(\cS^*_\infty[F;Y],L)$ denotes Lagrangian field
theory.

Cohomology of the Grassmann-graded variational bicomplex has been
obtained.$^{8,10}$ Let us mention the following relevant results.

\begin{theo} \mar{811} \label{811}
Cohomology of the variational complex (\ref{000}) equals the de
Rham cohomology of a fiber bundle $Y$.
\end{theo}

In particular, any odd element of this complex possesses trivial
cohomology.

\begin{theo} \label{g103} \mar{g103}
Given a graded Lagrangian $L$, there is the decomposition
\mar{g99,'}\ben
&& dL=\dl L - d_H\Xi_L,
\qquad \Xi\in \cS^{n-1}_\infty[F;Y], \label{g99}\\
&& \Xi_L=L+\op\sum_{s=0} \th^A_{\nu_s\ldots\nu_1}\w
F^{\la\nu_s\ldots\nu_1}_A\om_\la, \qquad \om_\la=\dr_\la\rfloor\om, \label{g99'}\\
&& F_A^{\nu_k\ldots\nu_1}= \dr_A^{\nu_k\ldots\nu_1}\cL-d_\la
F_A^{\la\nu_k\ldots\nu_1} +\psi_A^{\nu_k\ldots\nu_1},\qquad
k=1,2,\ldots,\nonumber
\een
where local graded functions $\psi$ obey the relations
\be
\psi^\nu_A=0,\qquad \psi_A^{(\nu_k\nu_{k-1})\ldots\nu_1}=0.
\ee
\end{theo}

The form $\Xi_L$ (\ref{g99'}) provides a global Lepage equivalent
of a graded Lagrangian $L$. In particular, one can locally choose
$\Xi_L$ (\ref{g99'}) where all functions $\psi$ vanish.

The corollaries of Theorem \ref{g103} are the first variational
formula (\ref{g107}) and the first Noether theorem (Theorem
\ref{j45}).

\bigskip
\bigskip

\noindent {\bf III. THE FIRST NOETHER THEOREM}
\bigskip

In order to treat symmetries of Lagrangian field theory
$(\cS^*_\infty[F;Y],L)$ in a very general setting, we consider
graded derivations of the $\Bbb R$-ring
$\cS^0_\infty[F;Y]$.$^{5,8}$ They take the form
\mar{gg3}\beq
 \vt=\vt^\la\dr_\la + \op\sum_{0\leq|\La|} \vt_\La^A\dr^\La_A,
 \qquad \dr^\La_A(s_\Si^B)=\dr^\La_A\rfloor
ds_\Si^B=\dl_A^B\dl^\La_\Si. \label{gg3}
\eeq
Any graded derivation $\vt$ (\ref{gg3}) yields the Lie derivative
\be
&& \bL_\vt\f=\vt\rfloor d\f+ d(\vt\rfloor\f), \qquad \f\in
\cS^*_\infty[F;Y],\\
&& \bL_\vt(\f\w\si)=\bL_\vt(\f)\w\si
+(-1)^{[\vt][\f]}\f\w\bL_\vt(\si),
\ee
of the DGA $\cS^*_\infty[F;Y]$.

A graded derivation $\vt$ (\ref{gg3}) is called contact if the Lie
derivative $\bL_\vt$ preserves the ideal of contact graded forms
of the DGA $\cS^*_\infty[F;Y]$. Any contact graded derivation
admits the decomposition
\mar{g105}\beq
\vt=\up_H+\up_V=\up^\la d_\la + [\up^A\dr_A +\op\sum_{|\La|>0}
d_\La(\up^A-s^A_\m\up^\m)\dr_A^\La] \label{g105}
\eeq
into the horizontal and vertical parts $\up_H$ and $\up_V$. A
glance at the expression (\ref{g105}) shows that a contact graded
derivation $\vt$ is an infinite order jet prolongation of its
restriction
\mar{jj15}\beq
\up=\up^\la\dr_\la +\up^A\dr_A \label{jj15}
\eeq
to the graded commutative ring $S^0[F;Y]$. One calls $\up$
(\ref{jj15}) the generalized graded vector field. It is a graded
vector field if its components $\up^\la$, $\up^A$ are independent
of jets $s^A_\La$. Note that generalized symmetries of Lagrangian
systems have been intensively studied.$^{5,11-13}$

Given a contact graded derivation (\ref{g105}), a corollary of the
decomposition (\ref{g99}) is the above mentioned first variational
formula
\mar{g107}\beq
\bL_\vt L= \up_V\rfloor\dl L +d_H(h_0(\vt\rfloor \Xi_L)) + d_V
(\up_H\rfloor\om)\cL, \label{g107}
\eeq
where $\Xi_L$ is the Lepage equivalent (\ref{g99'}) of $L$.

Given a Lagrangian $L$ (\ref{0709}), a contact graded derivation
$\vt$ (\ref{g105}) is said to be its variational symmetry
(strictly speaking a variational generalized supersymmetry) if the
Lie derivative $\bL_\vt L$ is $d_H$-exact, i.e.
\mar{001}\beq
\bL_\vt L=d_H\si. \label{001}
\eeq
A variational symmetry $\vt$ of a Lagrangian $L$ is called its
exact symmetry if the Lie derivative $\bL_\vt L$ vanishes.

An immediate corollary of the first variational formula
(\ref{g107}) is the following first Noether theorem.

\begin{theo} \label{j45} \mar{j45} If a contact graded derivation $\vt$
(\ref{g105}) is a variational symmetry (\ref{001}) of a Lagrangian
$L$, the first variational formula (\ref{g107}) restricted to
Ker$\,\dl L$ leads to the weak conservation law
\mar{35f2}\beq
d_H(\si-h_0(\vt\rfloor\Xi_L))\ap 0 \label{35f2}
\eeq
of the current
\mar{09200}\beq
\cJ_\vt=\cJ^\m_\vt\om_\m=\si- h_0(\vt\rfloor\Xi_L). \label{09200}
\eeq
\end{theo}

Obviously, the conserved current (\ref{09200}) is defined up to a
$d_H$-closed horizontal $(n-1)$-form
\mar{002}\beq
U=\frac12 U^{\nu\m}\om_{\nu\m}, \qquad
\om_{\nu\m}=\dr_\nu\rfloor\om_\m, \label{002}
\eeq
called the superpotential.

\begin{lem} \mar{35l10} \label{35l10}
A glance at the expression (\ref{g107}) shows the following.$^5$

(i) A contact graded derivation $\vt$ is a variational symmetry
only if the generalized vector field $\up$ (\ref{jj15}) is
projected onto $X$, i.e., $\up^\la\dr_\la$ is a vector field on
$X$.

(ii) A contact graded derivation $\vt$ is a variational symmetry
iff its vertical part $\up_V$ is well.

(iii) Any projectable contact graded derivation is a variational
symmetry of a variationally trivial Lagrangian.

(iv) A contact graded derivation $\vt$  is a variational symmetry
iff the graded density $\up_V\rfloor \dl L$ is $d_H$-exact.
\end{lem}

Variational symmetries of a Lagrangian $L$ constitute a real
vector space $\cG_L$.  By virtue of item (iii) of Lemma
\ref{35l10}, the Lie superbracket
\be
\bL_{[\vt,\vt']}=[\bL_\vt,\bL_{\vt'}]
\ee
of variational symmetries is a variational symmetry. Consequently,
the vector space $\cG_L$ of variational symmetries is a real Lie
superalgebra.

By virtue of item (ii) of Lemma \ref{35l10}, we further restrict
our consideration to vertical contact graded derivations
\mar{0672}\beq
\vt=\up^A\dr_A + \op\sum_{0<|\La|} d_\La\up^A\dr_A^\La.
\label{0672}
\eeq

A graded derivation $\vt$ (\ref{0672}) is called nilpotent if
$\bL_\vt(\bL_\vt\f)=0$ for any horizontal form $\f\in
\cS^{0,*}_\infty[F;Y]$. One can show that $\vt$ (\ref{0672}) is
nilpotent only if it is odd and iff $\vt(\up)=0$.$^5$

For the sake of brevity, the common symbol $\up$ further stands
for a generalized graded vector field $\up=\up^A\dr_A$, the
vertical contact graded derivation $\vt$ (\ref{0672}) determined
by $\up$, and the Lie derivative $\bL_\vt$. We agree to call $\up$
the graded derivation of the DGA $\cS^*_\infty[F;Y]$. The right
graded derivations $\op\up^\lto ={\op\dr^\lto}_A\up^A$ of
$\cS^*_\infty[F;Y]$ also are considered.

\bigskip
\bigskip

\noindent {\bf IV. GAUGE SYMMETRIES}
\bigskip

Without a loss of generality, let a Lagrangian $L$ be even. To
describe Noether identities of Lagrangian field theory
$(\cS^*_\infty[F;Y],L)$, let us introduce the following notation.
Given a vector bundle $E\to X$, we call
\be
\ol E=E^*\ot\op\w^n T^*X
\ee
the density-dual of $E$. The density dual of a graded vector
bundle $E=E^0\oplus E^1$ is $\ol E=\ol E^1\oplus \ol E^0$. Given a
graded vector bundle $E=E^0\oplus E^1$ over $Y$, we consider the
composite bundle $E\to E^0\to X$ and denote
$\cP^*_\infty[E;Y]=\cS^*_\infty[E;E^0]$. Let $VF$ be the vertical
tangent bundle of $F\to X$, the density-dual of the vector bundle
$VF\to F$ is
\be
\ol{VF}=V^*F\op\ot_F\op\w^n T^*X.
\ee

Let us enlarge $\cS^*_\infty[F;Y]$ to the DGA
$\cP^*_\infty[\ol{VF};Y]$ possessing the local basis $(s^A, \ol
s_A)$, $[\ol s_A]=([A]+1){\rm mod}\,2$. Its elements $\ol s_A$ are
called antifields. The DGA $\cP^*_\infty[\ol{VF};Y]$ is endowed
with the odd right graded derivation $\ol\dl=\rdr^A \cE_A$, where
$\cE_A$ are the variational derivatives (\ref{0709'}). This graded
derivation is obviously nilpotent. Then we have the chain complex
\mar{v042}\beq
0\lto \im\ol\dl \llr^{\ol\dl} \cP^{0,n}_\infty[\ol{VF};Y]_1
\llr^{\ol\dl} \cP^{0,n}_\infty[\ol{VF};Y]_2 \label{v042}
\eeq
of graded densities of antifield number $\leq 2$. Its one-cycles
\mar{0712}\beq
\ol\dl \Phi=0, \qquad \Phi= \op\sum_{0\leq|\La|} \Phi^{A,\La}\ol
s_{\La A} d^nx \in \cP^{0,n}_\infty[\ol{VF};Y]_1,\label{0712}
\eeq
define Noether identities of Lagrangian field theory
$(\cS^*_\infty[F;Y],L)$. In particular, one-chains $\Phi \in
\cP^{0,n}_\infty[\ol{VF};Y]_1$ are necessarily Noether identities
if they are boundaries. Therefore, these Noether identities are
called trivial. Accordingly, non-trivial Noether identities modulo
the trivial ones are associated to elements of the first homology
$H_1(\ol\dl)$ of the complex (\ref{v042}).$^{6,8}$

Let us assume that the homology $H_1(\ol \dl)$ is finitely
generated. Namely, there exists a projective $C^\infty(X)$-module
$\cC\subset H_1(\ol \dl)$ of finite rank possessing the local
basis $\{\Delta_r\}$ such that any element $\Phi\in H_1(\ol \dl)$
factorizes as
\mar{xx2}\beq
\Phi= \op\sum_{0\leq|\Xi|} G^{r,\Xi} d_\Xi \Delta_r d^nx, \qquad
\Delta_r=\op\sum_{0\leq|\La|} \Delta_r^{A,\La}\ol s_{\La A},\qquad
G^{r,\Xi},\Delta_r^{A,\La}\in \cS^0_\infty[F;Y], \label{xx2}
\eeq
through elements of $\cC$. Thus, all non-trivial Noether
identities (\ref{0712}) result from the Noether identities
\mar{v64}\beq
\ol\dl\Delta_r= \op\sum_{0\leq|\La|} \Delta_r^{A,\La} d_\La
\cE_A=0, \label{v64}
\eeq
called the complete Noether identities. By virtue of the
generalized Serre--Swan theorem,$^8$ the module $\cC$ is
isomorphic to the $C^\infty(X)$-module of sections of the
density-dual $\ol E$ of some graded vector bundle $E\to X$.

We define a non-trivial gauge symmetry of Lagrangian field theory
$(\cS^*_\infty[F;Y],L)$ as that associated to the Noether
identities (\ref{v64}) by means of the inverse second Noether
theorem.$^{6-8}$

Let us enlarge the DGA $\cP^*_\infty[\ol{VF};Y]$ to the DGA
$\cP^*_\infty[\ol{VF}\op\oplus_Y E;Y]$ possessing the local basis
$(s^A, \ol s_A, c^r)$. Its elements $c^r$ of Grassmann parity
$[c_r]=[\Delta_r]$ are called the ghosts. The graded derivation
$\ol \dl$ is naturally prolonged to the DGA
$\cP^*_\infty[\ol{VF}\op\oplus_Y E;Y]$. Let us extend an original
Lagrangian $L$ to the even Lagrangian
\mar{w8}\beq
L_e=L + c^r\Delta_r\om\in \cP^{0,n}_\infty[\ol{VF}\op\oplus_Y
E;Y]. \label{w8}
\eeq
It is readily observed that, by virtue of the Noether identities
(\ref{v64}), the graded derivation $\ol\dl$ is an exact symmetry
of $L_e$ (\ref{w8}). It follows from item (iv) of Lemma
\ref{35l10} that
\mar{w19}\beq
\frac{\op\dl^\lto (c^r\Delta_r)}{\dl \ol s_A}\cE_A \om
=\op\sum_{0\leq|\La|}
(-1)^{|\La|}d_\La(c^r\Delta_r^{A,\La})\cE_A\om= u^A\cE_A \om =
d_H\si. \label{w19}
\eeq
Then by the same reason, the odd graded derivation
\mar{w33}\beq
u= u^A\frac{\dr}{\dr s^A}, \qquad \qquad u^A =\op\sum_{0\leq|\La|}
c^r_\La\eta(\Delta^A_r)^\La, \label{w33}
\eeq
of $\cP^*_\infty[\ol{VF};Y]$ is a variational symmetry of an
original Lagrangian $L$.

A glance at the expression (\ref{w33}) shows that the variational
symmetry $u$ is a linear differential operator on the
$C^\infty(X)$-module $\cC$ of ghosts with values in the real space
$\cG_L$ of variational symmetries. It is called the gauge symmetry
of a Lagrangian $L$ which is associated to the complete
non-trivial Noether identities (\ref{v64}).

This association is unique due to the following. The variational
derivative of the equality (\ref{w19}) with respect to ghosts
$c^r$ leads to the equalities
\be
\dl_r(u^A\cE_A
\om)=\op\sum_{0\leq|\La|}(-1)^{|\La|}d_\La(\eta(\Delta^A_r)^\La\cE_A)=
\op\sum_{0\leq|\La|}\eta(\eta(\Delta^A_r))^\La d_\La\cE_A=
\op\sum_{0\leq|\La|}\Delta_r^{A,\La} d_\La\cE_A=0
\ee
which reproduce the complete non-trivial Noether identities
(\ref{v64}).

Moreover, the gauge symmetry $u$ (\ref{w33}) is complete in the
following sense. Let
\be
\op\sum_{0\leq|\Xi|} C^RG^{r,\Xi}_R d_\Xi \Delta_r \om
\ee
be some projective $C^\infty(X)$-module of finite rank of
non-trivial Noether identities parameterized by the corresponding
ghosts $C^R$. A direct computation shows that the graded
derivation
\be
d_\La(\op\sum_{0\leq|\Xi|}\eta(G^r_R)^\Xi
C^R_\Xi)u_r^{A,\La}\frac{\dr}{\dr s^A}
\ee
is a variational symmetry of a Lagrangian $L$ and, consequently,
its gauge symmetry parameterized by ghosts $C^R$.$^{7,8}$ It
factorizes through the gauge symmetry (\ref{w33}) by putting
ghosts
\be
c^r= \op\sum_{0\leq|\Xi|}\eta(G^r_R)^\Xi C^R_\Xi.
\ee

Thus, we come to the following second Noether theorem.

\begin{theo} \mar{s4} \label{s4}
The odd graded derivation $u$ (\ref{w33}) is a complete
non-trivial gauge symmetry of a Lagrangian $L$ associated to the
complete non-trivial Noether identities (\ref{v64}).
\end{theo}

\bigskip
\bigskip

\noindent {\bf V. GAUGE CONSERVATION LAWS}
\bigskip

Being a variational symmetry, the gauge symmetry  $u$ (\ref{w33})
defines the weak conservation law (\ref{35f2}). The peculiarity of
this conservation law is that the conserved current $\cJ_u$
(\ref{09200}) is reduced to a superpotential as follows.

\begin{theo} \mar{supp} \label{supp} If $u$ (\ref{w33}) is a gauge
symmetry of a Lagrangian $L$, the corresponding conserved current
$\cJ_u$ (\ref{09200}) takes the form
\mar{005}\beq
\cJ_u=W +d_HU=(W^\m + d_\nu U^{\nu\m})\om_\m, \label{005}
\eeq
where the form $W$ is $\ol\dl$-exact (i.e. it vanishes on-shell)
and $U$ is a superpotential (\ref{002}).
\end{theo}

\begin{proof}
Let the gauge symmetry $u$ (\ref{w33}) be at most of jet order $N$
in ghosts. Then the conserved current $\cJ_u$ (\ref{005}) is
decomposed into the sum
\mar{g2g}\ben
&& \cJ_u^\m= J^{\m\m_1\ldots\m_M}_rc^r_{\m_1\ldots\m_M} +
\op\sum_{1<k< M} J^{\m\m_k\ldots\m_M}_rc^r_{\m_k\ldots\m_M} + \label{g2g}\\
&& \qquad J^{\m\m_M}_rc^r_{\m_M} +J^\m_rc^r +J^\m, \qquad N\leq
M,\nonumber
\een
and the first variational formula (\ref{g107}) takes the form
\be
&& 0=\left[ \op\sum_{k=1}^N
u_V^i{}_r^{\m_k\ldots\m_N}c^r_{\m_k\ldots\m_N}
+u_V^i{}_rc^r\right]\cE_i -\\
&& \qquad d_\m\left(\op\sum_{k=1}^M
J^{\m\m_k\ldots\m_M}_rc^r_{\m_k\ldots\m_M} +J^\m_rc^r
+J^\m\right).
\ee
This equality provides the following set of equalities for each
$c^r_{\m\m_1\ldots\m_M}$, $c^r_{\m_k\ldots\m_M}$
$(k=1,\ldots,M-N-1)$, $c^r_{\m_k\ldots\m_N}$ $(k=1,\ldots,N-1)$,
$c^r_\m$ and $c^r$:
\mar{g4g,-6}\ben
&& 0=J^{(\m\m_1)\ldots\m_M}_r, \label{g4g}\\
&& 0=J^{(\m_k\m_{k+1})\ldots\m_M}_r +d_\nu
J^{\nu\m_k\ldots\m_M}_r, \qquad 1\leq k<M-N,
\label{g4g'}\\
&& 0=u_V^i{}_r^{\m_k\ldots\m_N}\cE_i-
J^{(\m_k\m_{k+1})\ldots\m_N}_r
-d_\nu J^{\nu\m_k\ldots\m_N}_r,\qquad 1\leq k<N, \label{g5g}\\
&& 0= u_V^i{}_r^\m\cE_i - J^\m_r - d_\nu J^{\nu\m}_r, \label{g6g}
\een
where $(\m\nu)$ means symmetrization of indices in accordance with
the splitting
\be
J^{\m_k\m_{k+1}\ldots\m_N}_r=J^{(\m_k\m_{k+1})\ldots\m_N}_r+
J^{[\m_k\m_{k+1}]\ldots\m_N}_r.
\ee
We also have the equalities
\mar{g7g,'}\ben
&& 0= u_V^i{}_r\cE_i - d_\m J^\m_r, \label{g7g}\\
&& 0=d_\m J^\m. \label{g7g'}
\een
With the equalities (\ref{g4g}) -- (\ref{g6g}), the decomposition
(\ref{g2g}) takes the form
\be
&& \cJ_u^\m= J^{[\m\m_1]\ldots\m_M}_rc^r_{\m_1\ldots\m_M} +\\
&& \qquad \op\sum_{1< k\leq M-N} [(J^{[\m\m_k]\ldots\m_M}_r -
d_\nu
J^{\nu\m\m_k\ldots\m_M}_r)c^r_{\m_k\ldots\m_M}]+ \\
&& \qquad \op\sum_{1<k< N}[(u_V^i{}_r^{\m\m_k\ldots\m_N}\cE_i -
d_\nu J^{\nu\m\m_k\ldots\m_N}_r +
 J^{[\m\m_k]\ldots\m_N}_r)c^r_{\m_k\ldots\m_N}]+\\
&& \qquad (u_V^i{}_r^{\m\m_N}\cE_i -d_\nu J^{\nu\m\m_N}_r +
J^{[\m\m_N]}_r)c^r_{\m_N} + (u_V^i{}_r^\m\cE_i - d_\nu
J^{\nu\m}_r)c^r + J^\m.
\ee
A direct computation
\be
&& \cJ_u^\m=
d_\nu(J^{[\m\nu]\m_2\ldots\m_M}_rc^r_{\m_2\ldots\m_M}) -
d_\nu J^{[\m\nu]\m_2\ldots\m_M}_rc^r_{\m_2\ldots\m_M}+\\
&&  \qquad \op\sum_{1< k\leq M-N}
[d_\nu(J^{[\m\nu]\m_{k+1}\ldots\m_M}_rc^r_{\m_{k+1}\ldots\m_M})
-\\
&&\qquad d_\nu
J^{[\m\nu]\m_{k+1}\ldots\m_M}_rc^r_{\m_{k+1}\ldots\m_M}- d_\nu
J^{\nu\m\m_k\ldots\m_M}_rc^r_{\m_k\ldots\m_M}]+\\
&& \qquad \op\sum_{1<k<N} [(u_V^i{}_r^{\m\m_k\ldots\m_N}\cE_i
- d_\nu J^{\nu\m\m_k\ldots\m_N}_r)c^r_{\m_k\ldots\m_N} +\\
&&  \qquad
d_\nu(J^{[\m\nu]\m_{k+1}\ldots\m_N}_rc^r_{\m_{k+1}\ldots\m_N})
-d_\nu J^{[\m\nu]\m_{k+1}\ldots\m_N}_rc^r_{\m_{k+1}\ldots\m_N}]+\\
&& \qquad [(u_V^i{}_r^{\m\m_N}\cE_i - d_\nu
J^{\nu\m\m_N}_r)c^r_{\m_N} + d_\nu (J^{[\m\nu]}_rc^r)
- d_\nu J^{[\m\nu]}_rc^r] +\\
&& \qquad (u_V^i{}_r^\m\cE_i - d_\nu J^{\nu\m}_r)c^r + J^\m\\
&& = d_\nu(J^{[\m\nu]\m_2\ldots\m_M}_rc^r_{\m_2\ldots\m_M})+\\
&&  \qquad \op\sum_{1< k\leq M-N}
[d_\nu(J^{[\m\nu]\m_{k+1}\ldots\m_M}_rc^r_{\m_{k+1}\ldots\m_M}) -
d_\nu
J^{(\nu\m)\m_k\ldots\m_M}_rc^r_{\m_k\ldots\m_M}]+\\
&& \qquad \op\sum_{1<k<N} [(u_V^i{}_r^{\m\m_k\ldots\m_N}\cE_i
- d_\nu J^{(\nu\m)\m_k\ldots\m_N}_r)c^r_{\m_k\ldots\m_N} +\\
&&  \qquad d_\nu(J^{[\m\nu]\m_{k+1}\ldots\m_N}_rc^r_{\m_{k+1}\ldots\m_N})]+\\
&& \qquad [(u_V^i{}_r^{\m\m_N}\cE_i - d_\nu
J^{(\nu\m)\m_N}_r)c^r_{\m_N} + d_\nu (J^{[\m\nu]}_rc^r)]
+(u_V^i{}_r^\m\cE_i - d_\nu J^{(\nu\m)}_r)c^r +J^\m
\ee
leads to the expression
\mar{g8g}\ben
&& \cJ_u^\m=\left(\op\sum_{1<k\leq N}u_V^i{}_r^{\m\m_k\ldots\m_N}
c^r_{\m_k\ldots\m_N}+ u_V^i{}_r^\m c^r\right)\cE_i -\label{g8g}\\
&& \qquad \left(\op\sum_{1<k\leq M}d_\nu
J^{(\nu\m)\m_k\ldots\m_M}c^r_{\m_k\ldots\m_M}+ d_\nu
J^{(\nu\m)}_rc^r\right)- \nonumber\\
&& \qquad d_\nu\left(\op\sum_{1<k\leq
M}J^{[\nu\m]\m_k\ldots\m_M}c^r_{\m_k\ldots\m_M} +
J^{[\nu\m]}_rc^r\right) +J^\m.\nonumber
\een
The first summand of this expression vanishes on-shell. Its second
one contains the terms $d_\nu J^{(\nu\m_k)\m_{k+1}\ldots\m_M}$,
$k=1,\ldots, M$. By virtue of the equalities (\ref{g4g'}) --
(\ref{g5g}), every $d_\nu J^{(\nu\m_k)\m_{k+1}\ldots\m_M}$ is
expressed into the terms vanishing on-shell and the term $d_\nu
J^{(\nu\m_{k-1})\m_k\ldots\m_M}$. Iterating the procedure and
bearing in mind the equality (\ref{g4g}), one can easily show that
the second summand of the expression (\ref{g8g}) also vanishes
on-shell. Finally, the condition (\ref{g7g'}) means that the odd
$(n-1)$-form $J^\m\om_\m$ is $d_H$-closed and, consequently, it is
$d_H$-exact in accordance with Theorem \ref{811}. Thus, the
current $\cJ_u$ takes the form (\ref{005}).
\end{proof}


\begin{thebibliography}{ddd}

\bibitem{julia} B.Julia and S. Silva, Currents and superpotentials
in classical gauge inveriant theories. Local results with
applications to perfect dluids and General Relativity, {\it Class.
Quant. Grav.} {\bf 15} (1998) 2173.

\bibitem{got92} M.Gotay and J.Marsden, Stress-energy-momentum
tensors and the Belinfante--Rosenfeld formula, {\it Contemp.
Math.} {\bf 132} (1992) 367.

\bibitem{fat94} L.Fatibene, M.Ferraris and M.Francaviglia,  (1994).
N\"other formalism for conserved quantities in classical gauge
field theories. {\it J. Math. Phys.} {\bf 35} (1994) 1644.

\bibitem{book} G.Giachetta, L.Mangiarotti and G.Sardanashvily,
{\it New Lagrangianm and Hamiltonian Methods in Field Theory}
(World Scientific, Singapore, 1997).

\bibitem{cmp04} G.Giachetta, L.Mangiarotti, and G.Sardanashvily,
Lagrangian supersymmetries depending on derivatives. Global
analysis and cohomology. {\it Commun. Math. Phys.} {\bf 259}
(2005) 103; {\it E-print arXiv}: hep-th/0407185.

\bibitem{lmp08} D.Bashkirov, G.Giachetta, L.Mangiarotti and
G.Sardanashvily, The KT-BRST complex of a degenerate Lagrangian
system, {\it Lett. Math. Phys.} {\bf 83} (2008) 237; {\it E-print
arXiv:} math-ph/0702097.

\bibitem{jmp09} G.Giachetta, L.Mangiarotti and
G.Sardanashvily, On the notion of gauge symmetries of generic
Lagrangian field theory, {\it J. Math. Phys.} {\bf 50} (2009)
012903; {\it E-print arXiv:} 0807.3003.


\bibitem{book09} G.Giachetta, L.Mangiarotti and
G.Sardanashvily, {\it Advanced Classical Field Theory} (World
Scientific, Singapore, 2009).

\bibitem{barn} G.Barnich, F.Brandt and M.Henneaux, Local
BRST cohomology in gauge theories, {\it Phys. Rep.} {\bf 338}
(2000) 439.

\bibitem{ijgmmp07} G.Sardanashvily, Graded infinite order jet manifolds,
{\it Int. J. Geom. Methods Mod. Phys.} {\bf 4} (2007) 1335; {\it
E-print arXiv:} 0708.2434.

\bibitem{olv} P. Olver, {\it Applications of Lie Groups to
Differential Equations} (Springer, Berlin, 1986).

\bibitem{fat} L.Fatibene, M.Ferraris, M.Francaviglia
and R.McLenaghan, Generalized symmetries in mechanics and field
theories, {\it J. Math.. Phys.} {\bf 43} (2002) 3147.

\bibitem{bry} R.Bryant, P.Griffiths and D.Grossman,
{\it Exterior Differential Systems and Euler--Lagrange Partial
Differential Equations} (Univ. of Chicago Press, Chicago, IL,
2003).



\end{thebibliography}
\end{document}